\documentclass[11pt,letterpaper]{article}
\usepackage[T1]{fontenc}
\usepackage[latin9]{inputenc}
\usepackage{textcomp}
\usepackage{amsmath}
\usepackage{amssymb}
\usepackage{esint}

\makeatletter

\pdfpageheight\paperheight
\pdfpagewidth\paperwidth

\pdfoutput=1

\usepackage[T1]{fontenc}
\usepackage[latin9]{inputenc}
\usepackage[a4paper]{geometry}
\usepackage[active]{srcltx}
\usepackage{amsmath}
\usepackage{amssymb}
\usepackage{esint}
\usepackage{ulem}

\usepackage{xcolor}

\makeatletter

\usepackage{textcomp}

\pdfoutput=1 

\usepackage{jheppub}

\usepackage{etoolbox}
    
   \patchcmd{\maketitle}{\@fpheader}{}{}{}

\newcommand*\xbar[1]{%
  \hbox{%
    \vbox{%
      \hrule height 0.5pt 
      \kern0.3ex
      \hbox{%
        \kern-0.0em
        \ensuremath{#1}%
        \kern-0.0em
      }%
    }%
  }%
}

\usepackage{amsfonts}

\usepackage{bbm}

\newcommand{\be}{\begin{equation}}
\newcommand{\ee}{\end{equation}}
\newcommand{\bea}{\begin{eqnarray}}
\newcommand{\eea}{\end{eqnarray}}

\title{$p$-form electrodynamics as edge modes of a topological field theory}

\author[a]{Oscar Fuentealba,}
\author[b,c]{Ricardo Troncoso}

\affiliation[a]{Universit\'e Libre de Bruxelles and International Solvay Institutes, ULB-Campus Plaine CP231, B-1050 Brussels, Belgium}
\affiliation[b]{Centro de Estudios Cient\'{\i}ficos (CECs),  Av. Arturo Prat 514, Valdivia, Chile}
\affiliation[c]{Facultad de Ingenier\'{i}a, Arquitectura y Dise\~{n}o, Universidad San Sebasti\'an, sede Valdivia, General Lagos 1163, Valdivia 5110693, Chile}

\emailAdd{oscar.fuentealba@ulb.be}
\emailAdd{ricardo.troncoso@uss.cl}

\preprint{}

\abstract{$p$-form electrodynamics in $d\geq 2$ dimensions is shown to emerge as the edge modes of a topological field theory with a precise set of boundary conditions, through the Hamiltonian reduction of its action. Electric and magnetic charges correspond to Noether ones in the topological field theory. For chiral $p$-forms, the topological action can be consistently truncated, so that the Henneaux-Teitelboim action is recovered from a pure Chern-Simons theory, with a manifestly covariant stress-energy tensor at the boundary. Topologically massive $p$-form electrodynamics as well as axion couplings are also shown to be described through this mechanism by considering suitable (self-)interaction terms in the topological theory.} 

\makeatother

\begin{document}
\maketitle 

\flushbottom

\newpage{}

\section{Introduction}

The massless Klein-Gordon field, Maxwell electrodynamics and the Kalb-Ramond
field are well-known to be described in a unified way through $p$-form
electrodynamics with $p=0,1,2$, respectively (see e.g., \cite{Kalb:1974yc,Nambu:1975ba,Curtright:1980yj,Sezgin:1980tp,Siegel:1980jj,Townsend:1979yv,Orland:1981ku,Schwarz:1983wa,Freedman:1980us,Rohm:1985jv,Teitelboim:1985ya,Teitelboim:1985yc,Henneaux:1986ht,Hsieh:2020jpj}).
$p$-form electrodynamics also plays a very relevant role in the description
of supergravity and string theory in diverse dimensions \cite{VanNieuwenhuizen:1981ae,Freedman:2012zz,Green:1987sp,Polchinski:1998rq}.

One of the main aims of our work is showing that $p$-form electrodynamics
in $d\geq2$ dimensions emerges as the edge modes of a topological
field theory of BF-type once endowed with a very precise set of boundary
conditions, defined in Section \ref{section2}, that requires the
presence of a metric only at the boundary. The form of the stress-energy
tensor at the boundary naturally suggests the link with $p$-form
electrodynamics, which is explicitly performed through the Hamiltonian
reduction of the topological action in Section \ref{section3}. It
is worth stressing that electric and magnetic charges can then be
seen as Noetherian ones in the BF theory. The case of even $p$-forms
in $d=2p+2$ spacetime dimensions is discussed in Section \ref{section4},
where it is shown that the topological action can be consistently
truncated to a pure Chern-Simons theory devoid of boundary terms,
whose Hamiltonian reduction precisely yields the Henneaux-Teitelboim
action for chiral $p$-forms. We conclude in Section \ref{Ending-remarks},
where (self-)interactions of the topological field theory are considered,
which allows to reproduce topologically massive $p$-form electrodynamics,
extensions of it, as well as axion couplings as edge modes for the
same set of boundary conditions. 

\section{Topological field theory of BF-type \label{section2}}

Let us consider the action principle of an Abelian BF theory (see
e.g., \cite{Schwarz:1978cn,Schwarz:1979ae,Horowitz:1989ng,Blau:1989dh,Horowitz:1989km,Oda:1989tq,Birmingham:1991ty,Blau:1989bq,Wu:1990ci,Gegenberg:1993gd})
on a manifold $\Omega$ of $d+1$ dimensions, given by
\begin{equation}
I=\intop_{\Omega}B\wedge dC+\frak{B}\,,\label{eq:IBF}
\end{equation}
where $B$ and $C$ correspond to $p+1$ and $\left(d-p-1\right)$-forms,
respectively. The boundary term $\frak{B}$ is defined on $M=\partial\Omega$,
being generically required in order to have a well-defined variational
principle once the boundary conditions are specified. Its precise
form can be obtained as follows. The variation of \eqref{eq:IBF}
reads
\begin{equation}
\delta I=\intop_{\Omega}\left(\delta B\wedge dC+(-1)^{p}dB\wedge\delta C\right)-(-1)^{p}\intop_{M}B\wedge\delta C+\delta\frak{B}\,,\label{eq:dIBF}
\end{equation}
so that the bulk terms vanish when the field equations, $dB=0$ and
$dC=0$, hold. Thus, the action attains an extremum provided that
the variation of the boundary term is given by
\begin{align}
\delta\frak{B} & =(-1)^{p}\intop_{M}B\wedge\delta C\,,\label{eq:bt}
\end{align}
which requires a precise choice of boundary conditions to be integrated.

\subsection{Boundary conditions}

In order to specify our boundary conditions we assume the existence
of a metric structure at the boundary, so that $g_{\mu\nu}$ is defined
only at $M=\partial\Omega$. The boundary conditions are then defined
by choosing the $C$-field to be the Hodge dual the $B$-field at
the boundary, i.e.,\footnote{Note that $M$ is also assumed to be orientable, so that the Hodge
dual of an $r$-form $\omega$ is defined as $*\omega=\frac{\sqrt{-g}}{r!(d-r)!}\omega^{\mu_{1}\cdots\mu_{r}}\epsilon_{\mu_{1}\cdots\mu_{r}\nu_{r+1}\cdots\nu_{d}}dx^{\nu_{r+1}}\wedge\cdots\wedge dx^{\nu_{d}}$.}
\begin{equation}
\left(C-*B\right)\Big|_{M=\partial\Omega}=0\,.\label{eq:bc}
\end{equation}

The boundary condition \eqref{eq:bc} then allows to integrate the
variation of the boundary term $\delta\frak{B}$ in \eqref{eq:bt},
so that it is given by
\begin{align}
\frak{B} & =\frac{(-1)^{p}}{2}\intop_{M}B\wedge*B\,.\label{eq:IB}
\end{align}

In sum, the action principle
\begin{equation}
I=\intop_{\Omega}B\wedge dC+\frac{(-1)^{p}}{2}\intop_{M}B\wedge*B\,.\label{eq:IBFwell-def}
\end{equation}
becomes well-defined for our choice of boundary conditions in \eqref{eq:bc}.

\subsection{Stress-energy tensor at the boundary \label{Stress-energy p-forms}}

Note that the topological field theory under discussion has no local
notion of energy. Nevertheless, since the boundary conditions incorporate
a metric, it is possible to define a stress-energy tensor at the boundary
$M=\partial\Omega$ along the lines of Brown and York \cite{Brown:1992br},
given by
\begin{equation}
T_{\mu\nu}=-\frac{2}{\sqrt{-g}}\frac{\delta I}{\delta g^{\mu\nu}}\,,\label{eq:Tmunu}
\end{equation}
which for the action in \eqref{eq:IBFwell-def} reads
\begin{equation}
T_{\mu\nu}=\frac{\left(-1\right)^{(p+1)(d-p)}}{\left(p+1\right)!}\left(\left(p+1\right)B_{\mu\mu_{1}\cdots\mu_{p}}B_{\nu}^{\,\,\,\mu_{1}\cdots\mu_{p}}-\frac{1}{2}g_{\mu\nu}B_{\mu_{1}\cdots\mu_{p+1}}B^{\mu_{1}\cdots\mu_{p+1}}\right)\,.\label{eq:TmunuP-form}
\end{equation}

The explicit form of the stress-energy tensor in \eqref{eq:TmunuP-form}
then naturally suggests a link with $p$-form electrodynamics at the
boundary, which is discussed in the next section.

\section{$p$-form electrodynamics from Hamiltonian reduction \label{section3}}

The Hamiltonian reduction of the topological field theory described
by \eqref{eq:IBFwell-def} can be readily performed due to its simplicity.
Assuming the topology of $\Omega$ to be of the form $\mathbb{R}\times\Sigma$,
indices can be split in space and time so that the action reads
\begin{equation}
I={\displaystyle \frac{1}{\alpha}}\intop_{\Omega}dtd^{d}x\Big[\left(-1\right){}^{p}\epsilon^{0i_{1}\cdots i_{d}}\dot{B}_{i_{1}\cdots i_{p+1}}C_{i_{p+2}\cdots i_{d}}+B_{0i_{1}\cdots i_{p}}G_{C}^{i_{1}\cdots i_{p}}+C_{0i_{1}\cdots i_{d-p-2}}G_{B}^{i_{1}\cdots i_{d-p-2}}\Big]+\tilde{\mathcal{\frak{B}}}\,,\label{eq:Iham}
\end{equation}
with $\alpha=\left(p+1\right)!(d-p-1)!$, and the boundary term is
given by
\begin{equation}
\tilde{\frak{B}}=\frak{B}+\frac{\left(-1\right)^{p(d-p-1)}}{\left(p+1\right)!(d-p-2)!}\intop_{\Omega}dtd^{d}x\epsilon^{0i_{1}\cdots i_{d}}\partial_{i_{d}}\Big(C_{0i_{1}\cdots i_{d-p-2}}B_{i_{d-p-1}\cdots i_{d-1}}\Big)\,.
\end{equation}
Note that $B_{0i_{1}\cdots i_{p}}$ and $C_{0i_{1}\cdots i_{d-p-2}}$
stand for Lagrange multipliers, and hence the corresponding constraints
fulfilling\textbf{
\begin{align}
G_{C}^{i_{1}\cdots i_{p}} & :=(p+1)\epsilon^{0i_{1}\cdots i_{d}}\partial_{i_{p+1}}C_{i_{p+2}\cdots i_{d}}=0\,,\\
G_{B}^{i_{1}\cdots i_{d-p-2}} & :=-\left(-1\right)^{p(d-p-1)}(d-p-1)\epsilon^{0i_{1}\cdots i_{d}}\partial_{i_{d}}B_{i_{d-p-1}\cdots i_{d-1}}=0\,,
\end{align}
}are locally solved by
\begin{eqnarray}
B_{i_{1}\cdots i_{p+1}} & = & (p+1)\partial_{[i_{1}}A_{\cdots i_{p+1}]}\quad,\quad C_{i_{1}\cdots i_{d-p-1}}=\partial_{[i_{1}}\tilde{A}_{\cdots i_{d-p-1}]}\,.\label{eq:B sol}
\end{eqnarray}
Thus, replacing the solution of the constraints in \eqref{eq:B sol}
back into \eqref{eq:Iham}, after suitable integration by parts, the
full action reduces to a boundary term that can be written as
\begin{equation}
I=-\frac{(-1)^{p}(p+1)}{\alpha}\intop_{M}dtd^{d-1}x\epsilon^{\mu_{1}\cdots\mu_{d}}\partial_{\mu_{1}}A_{\mu_{2}\cdots\mu_{p+1}}C_{\mu_{p+2}\cdots\mu_{d}}+\frak{B}\,.\label{eq:I ham2}
\end{equation}
It is then useful to fix the gauge according to
\begin{equation}
B_{0i_{1}\cdots i_{p}}=(p+1)\partial_{[0}A_{i_{1}\cdots i_{p}]}\,,\label{eq:B0}
\end{equation}
so that $B=dA$. Besides, the boundary condition \eqref{eq:bc} allows
to trade the $C$-field by the Hodge dual of $B$, and hence, the
action \eqref{eq:I ham2} becomes that of $p$-form electrodynamics,
given by
\begin{equation}
I[A]=-\frac{(-1)^{p}}{2}\intop_{M}B\wedge*B\,,\label{eq:I-p-ed}
\end{equation}
where the dynamical field turns out to be the $p$-form $A$, whose
field strength is $B=dA$.

One interesting direct consequence of the equivalence of the topological
action \eqref{eq:IBFwell-def} with that of $p$-form electrodynamics
at the boundary, is that electric and magnetic charges can both be
seen to emerge from Noether ones in the topological theory. Indeed,
the suitably normalized conserved charges associated to the gauge
transformations of the topological action ($\delta B=d\lambda_{B}$,
$\delta C=d\lambda_{C}$) once evaluated at the boundary are given
by
\begin{equation}
Q_{B}\Big|_{\partial\Omega}=\intop C=\int*B\quad,\quad Q_{C}\Big|_{\partial\Omega}=\int B\,,
\end{equation}
corresponding to the electric and magnetic charges of $p$-form electrodynamics,
respectively.

\section{Consistent truncation: chiral $p$-forms from a pure Chern-Simons
theory \label{section4}}

Let us consider the case of even $p$-forms for $d=2p+2$, so that
one can perform the following change of basis in the fields of the
topological theory
\begin{equation}
B^{\pm}=\frac{1}{2}\left(B\pm C\right)\,.\label{eq:ChangeBasis}
\end{equation}
In terms of $B^{\pm}$, the boundary condition \eqref{eq:bc} then
reads
\begin{equation}
\left(B^{\pm}\mp*B^{\pm}\right)\Big|_{M=\partial\Omega}=0\,,\label{eq:Bpm}
\end{equation}
which amounts to require the fields to be (anti-)selfdual at the boundary,
and the topological action \eqref{eq:IBFwell-def} becomes just the
difference of two pure Chern-Simons forms, given by
\begin{equation}
I=I^{+}-I^{-}=\intop_{\Omega}B^{+}\wedge dB^{+}-\intop_{\Omega}B^{-}\wedge dB^{-}\,.\label{eq:ActionCS's}
\end{equation}
Note that the simple change of basis \eqref{eq:ChangeBasis} yields
to the action \eqref{eq:ActionCS's} that is devoid of boundary terms,
and leads to a well-defined variational principle for (anti-)chiral
fields at the boundary by construction.

It is worth highlighting that the action \eqref{eq:ActionCS's} can
be consistently truncated to describe single chiral fields at the
boundary, either for vanishing $B^{+}$or $B^{-}$. The link between
chiral $p$-forms and pure Chern-Simons theories has also been explored
in \cite{Belov:2006jd}.

\subsection{Chiral $p$-form action from Hamiltonian reduction}

The Hamiltonian reduction of the action 

\begin{equation}
I^{\pm}=\intop_{\Omega}B^{\pm}\wedge dB^{\pm}\,,\label{eq:ICSwell-def}
\end{equation}
for an odd $(p+1)$-form $B^{\pm}$ in $2p+3$ spacetime dimensions
that is (anti-)chiral at the boundary, is performed along the lines
of Section \ref{section3}. Splitting the indices in space and time,
the action \eqref{eq:ICSwell-def} can be written as
\begin{equation}
I^{\pm}=\frac{1}{\left(p+1\right)!^{2}}\intop_{\Omega}dtd^{d}x\epsilon^{0i_{1}\cdots i_{2p+2}}\Big[\dot{B}_{i_{1}\cdots i_{p+1}}^{\pm}B_{i_{p+2}\cdots i_{2p+2}}^{\pm}+2(p+1)B_{0i_{1}\cdots i_{p}}^{\pm}\partial_{i_{p+1}}B_{i_{p+2}\cdots i_{2p+2}}^{\pm}\Big]+\hat{\frak{B}}^{\pm}\,,\label{eq:I-CS-ham1}
\end{equation}
with a boundary term $\hat{\frak{B}}^{\pm}$ that reads
\begin{equation}
\hat{\frak{B}}^{\pm}=\frac{\left(p+1\right)}{\left(p+1\right)!^{2}}\intop_{\Omega}dtd^{d}x\epsilon^{0i_{1}\cdots i_{2p+2}}\partial_{i_{2p+2}}\Big(B_{0i_{1}\cdots i_{p}}^{\pm}B_{i_{p+1}\cdots i_{2p+1}}^{\pm}\Big)\,.
\end{equation}
The constraint associated to the Lagrange multiplier $B_{0i_{1}\cdots i_{p}}^{\pm}$in
\eqref{eq:I-CS-ham1} is then exactly solved as
\begin{equation}
B_{i_{1}\cdots i_{p+1}}^{\pm}=\partial_{[i_{1}}A_{\cdots i_{p+1}]}^{\pm}\,,\label{eq:Sol-constr-self}
\end{equation}
so that the the full action \eqref{eq:I-CS-ham1} reduces to a boundary
term given by
\begin{equation}
I^{\pm}=\frac{1}{\left(p+1\right)!^{2}}\intop_{M}dtd^{d-1}x\epsilon^{0a_{1}\cdots a_{2p+1}}\left(-\dot{A}_{a_{1}\cdots a_{p}}^{\pm}B_{a_{p+1}\cdots a_{2p+1}}^{\pm}+(p+1)B_{0a_{1}\cdots a_{p}}^{\pm}B_{a_{p+1}\cdots a_{2p+1}}^{\pm}\right)\,,\label{eq:I-CS-ham2}
\end{equation}
where latin indices $a_{1},\dots,a_{2p+1}$ stand for the spacelike
ones at $M=\partial\Omega$. The (anti-)self-duality condition \eqref{eq:Bpm}
fixes the Lagrange multiplier in terms of the field strength in \eqref{eq:Sol-constr-self},
according to
\begin{equation}
B_{0a_{1}\cdots a_{p}}^{\pm}=\pm\frac{1}{\left(p+1\right)!}\sqrt{-g}B_{\pm}^{\mu_{1}\cdots\mu_{p+1}}\epsilon_{\mu_{1}\cdots\mu_{p+1}0a_{1}\dots a_{p}}\,,
\end{equation}
and hence, the action \eqref{eq:I-CS-ham2}, reduces to the Henneaux-Teitelboim
action for (anti-)chiral $p$-forms \cite{Henneaux:1987hz,Henneaux:1988gg},
given by
\begin{equation}
I^{\pm}[A^{\pm}]=\mp\frac{p!}{\left(p+1\right)!^{2}}\intop_{\Omega}dtd^{d-1}x\left[\pm\mathcal{B}_{\pm}^{a_{1}\dots a_{p}}\dot{A}_{a_{1}\cdots a_{p}}^{\pm}-\left(N\mathcal{H}^{\pm}+N^{a}\mathcal{H}_{a}^{\pm}\right)\right]\,,\label{eq:HTaction}
\end{equation}
written in terms of the magnetic field

\begin{equation}
\mathcal{B}_{\pm}^{a_{1}\dots a_{p}}=\frac{1}{p!}\epsilon^{0a_{1}\dots a_{2p+1}}B_{a_{p+1}\dots a_{2p+1}}^{\pm}\,.
\end{equation}
Here, we have also made use of the ADM decomposition of the metric
\begin{equation}
ds^{2}=-N^{2}dt^{2}+\gamma_{ab}(dx^{a}+N^{a}dt)(dx^{b}+N^{b}dt)\,,
\end{equation}
so that $N$ and $N^{a}$ stand for the lapse and shift functions,
respectively, while the energy and momentum densities explicitly read
\begin{equation}
\mathcal{H}^{\pm}=\frac{1}{\sqrt{\gamma}}\mathcal{B}_{\pm}^{a_{1}\dots a_{p}}\mathcal{B}_{a_{1}\dots a_{p}}^{\pm},\label{eq:Energy}
\end{equation}
\begin{equation}
\mathcal{H}_{a}^{\pm}=\pm\frac{1}{p!}\epsilon_{\,\,\,a_{1}\dots a_{2p}a}^{0}\mathcal{B}_{\pm}^{a_{1}\dots a_{p}}\mathcal{B}_{\pm}^{a_{p+1}\dots a_{2p}}.\label{eq:Momentum}
\end{equation}

It is worth pointing out that although covariance is not manifest
in the action \eqref{eq:HTaction}, invariance under diffeomorphisms
that preserve the background metric holds by virtue of the fact that
the energy and momentum densities fulfill the Dirac-Schwinger algebra
\cite{Henneaux:1987hz,Henneaux:1988gg}.

\subsection{Manifestly covariant stress-energy tensor for chiral $p$-forms}

One of the advantages of obtaining the Henneaux-Teitelboim action
for chiral $p$-forms \eqref{eq:HTaction} as an edge mode of the
pure Chern-Simons action in \eqref{eq:ICSwell-def} is that a manifestly
covariant stress-energy tensor can be readily obtained as in Section
\ref{Stress-energy p-forms}. Indeed, the Brown-York stress-energy
tensor in \eqref{eq:Tmunu} evaluated at the boundary $M=\partial\Omega$
is given by
\begin{equation}
T_{\mu\nu}^{\pm}=\pm\frac{2}{p!}B_{\mu\mu_{1}\cdots\mu_{p}}^{\pm}B_{\nu}^{\pm\,\,\,\mu_{1}\cdots\mu_{p}}\,,\label{eq:Tmunu-self}
\end{equation}
being traceless, manifestly covariant and clearly conserved by virtue
of the Bianchi identity and the (anti-)self-dual boundary condition
\eqref{eq:Bpm}. Its components relate to the energy and momentum
densities by projecting along the normal vector to the spacelike hypersurface
$n_{\mu}=(N,0)$, so that
\begin{align}
T_{\perp\perp}^{\pm} & =T_{\pm}^{\mu\nu}n_{\mu}n_{\nu}=\pm\frac{2}{\left(p-1\right)^{2}p!\sqrt{\gamma}}\mathcal{H}^{\pm}\,,\\
T_{\perp a}^{\pm} & =T_{\pm\,\,a}^{\mu}n_{\mu}=\pm\frac{2}{\left(p-1\right)^{2}p!\sqrt{\gamma}}\mathcal{H}_{a}^{\pm}\,,
\end{align}
where $\mathcal{H}^{\pm}$ and $\mathcal{H}_{a}^{\pm}$ are given
by \eqref{eq:Energy} and \eqref{eq:Momentum}, respectively.

\section{Extensions and final remarks \label{Ending-remarks}}

Axion-like couplings between diverse $p$-forms as well as topologically
massive extensions can also be seen to emerge as edge modes of topological
field theories that include suitable (self-)interaction terms deforming
the gauge symmetries without breaking them. 

\subsection{Topologically massive $p$-form electrodynamics from a BF theory
with a ``cosmological term''}

One possibility is to endow the topological theory in \eqref{eq:IBFwell-def}
with a ``cosmological term'', extending that in \cite{Horowitz:1989ng}
for even $(p+1)$-forms in higher dimensions, so that our action in
\eqref{eq:IBFwell-def} is deformed as
\begin{equation}
I_{(\mu)}=I+\frac{\mu}{2}\intop_{\Omega}B^{m}\,,\label{eq:BF-cosmological}
\end{equation}
being clearly well-defined for the same boundary conditions in \eqref{eq:bc},
provided that $B$ stands for an even $(p+1)$-form in $d+1=m(p+1)$
dimensions. Note that the integer $m$ ranges as $2\leq m\leq(d+1)/2$,
and the gauge symmetries now become 
\begin{equation}
\delta B=d\lambda_{B}\quad,\quad\delta C=d\lambda_{C}-\frac{m(m-1)}{2}\mu B^{m-2}\lambda_{B}\,.
\end{equation}

The Hamiltonian reduction can then be carried out as in Section \ref{section3},
so that once the indices are split in space and time, the action is
given by \eqref{eq:Iham} with a deformed constraint $G_{C}^{i_{1}\cdots i_{p}}\rightarrow G_{(\mu)C}^{i_{1}\cdots i_{p}}$,
with

\textbf{
\begin{equation}
G_{(\mu)C}^{i_{1}\cdots i_{p}}=G_{C}^{i_{1}\cdots i_{p}}+\frac{(d-p-1)!m}{2p![(p+1)!]^{m-2}}\mu\left(B^{m-1}\right)^{i_{1}\cdots i_{p}}=0\,,
\end{equation}
}where $\left(B^{m-1}\right)^{i_{1}\cdots i_{p}}=\epsilon^{0i_{1}\cdots i_{p}i_{p+1}\cdots i_{d}}B_{i_{p+1}\cdots i_{2p+1}}\cdots B_{i_{d-p}\cdots i_{d}}$.
Thus, the constraints can also be locally solved, and once the solution
is replaced back into the action, fixing the gauge as in \eqref{eq:B0},
it reduces to a boundary term given by
\begin{equation}
I_{(\mu)}[A]=-\frac{(-1)^{p}}{2}\intop_{M}\left(B\wedge*B+\mu A\wedge B^{m-1}\right)\,,\label{eq:TME}
\end{equation}
where the $p$-form $A$ is the dynamical field with field strength
$B=dA$, precisely reproducing topologically massive $p$-form electrodynamics
for $m=2$, and extending it otherwise. In particular, for a standard
$U(1)$ gauge field ($p=1$), the original topologically massive electrodynamics
of Deser, Jackiw and Templeton \cite{Deser:1981wh} is recovered in
$d=3$ ($m=2$), while the graviphoton of five-dimensional supergravity
is obtained in $d=5$ ($m=3$) for a precise value of the deformation
parameter $\mu$. More possibilities arise in higher dimensions, as
in the case of $d=11$, where three different theories can be obtained,
for $p=1,3,5$ (with $m=6,3,2$, respectively). Note that the eleven-dimensional
supergravity $3$-form field \cite{Cremmer:1978km} is described by
\eqref{eq:TME}, where the value of $\mu$ becomes fixed by local
supersymmetry.

\subsection{Axion-like couplings from interacting topological theories}

A class of couplings between $p$-form fields for diverse values of
$p$ can also be described through the edge modes of a topological
theory of BF-type with suitable interaction terms. 

As a precise example let us consider a five-dimensional action of
the form
\begin{equation}
I_{(\lambda)}=\intop_{\Omega}\left(B_{[2]}\wedge dC_{[2]}-B_{[1]}\wedge dC_{[3]}+\frac{\lambda}{2}B_{[2]}^{2}\wedge B_{[1]}\right)-\frac{1}{2}\intop_{M}\left(B_{[2]}\wedge*B_{[2]}+B_{[1]}\wedge*B_{[1]}\right)\,,\label{eq:I-int-BF}
\end{equation}
which is clearly well-defined for boundary conditions as in \eqref{eq:bc},
regardless the value of the coupling $\lambda$. Following the same
lines as in the previous cases, the Hamiltonian reduction of \eqref{eq:I-int-BF}
reduces to a four-dimensional boundary term describing the axion coupling
of Maxwell electrodynamics with the massless Klein-Gordon field, given
by
\begin{equation}
I_{(\lambda)}[A,\phi]=\frac{1}{2}\intop_{M}\left(B_{[2]}\wedge*B_{[2]}+B_{[1]}\wedge*B_{[1]}+\lambda\phi B_{[2]}\wedge B_{[2]}\right)\,,
\end{equation}
where $B_{[2]}=dA$ and $B_{[1]}=d\phi$.

As a closing remark, it is certainly worth exploring how topological
invariants as the Ray-Singer torsion and the generalized linking number,
known to be deeply connected with topological field theories of BF-type
\cite{Schwarz:1978cn,Schwarz:1979ae,Horowitz:1989ng,Blau:1989dh,Horowitz:1989km,Oda:1989tq,Birmingham:1991ty,Blau:1989bq,Wu:1990ci,Gegenberg:1993gd},
reflect themselves in the context of $p$-form electrodynamics. \newline

\textbf{Note added}. This is a slightly updated version of our unpublished
preprint \cite{Fuentealba-Troncoso2012} that was presented in XVIII
Chilean Symposium of Physics during November 2012. Our results possess
some overlap with those recently reported in \cite{Evnin:2023cdf}.

\acknowledgments We would like to thank Claudio Bunster, Marcela
Cárdenas, Hernán A. González, Marc Henneaux, Javier Matulich, Alfredo
Pérez, Miguel Pino, David Tempo and Cédric Troessaert for many useful
discussions along the years. O.F. wishes to thank to the organizers
of the XVIII Chilean Symposium of Physics hosted by Universidad de
La Serena and Sociedad Chilena de Física (SOCHIFI), during November
2012, for the opportunity of presenting this work. This research has
been partially supported by ANID FONDECYT grants N° 1211226, 1220910,
1221624. The work of O.F. was partially supported by a Marina Solvay
Fellowship, FNRS-Belgium (conventions FRFC PDRT.1025.14 and IISN 4.4503.15),
as well as by funds from the Solvay Family.

\appendix

\end{document}